Research Paper

# MicroShare: Privacy-Preserved Medical Resource Sharing through MicroService Architecture

Yilong Yang[1,2], Quan Zu[2], Peng Liu[2], Defang Ouyang[1], Xiaoshan Li[2✉]

1. State Key Laboratory of Quality Research in Chinese Medicine, Institute of Chinese Medical Sciences (ICMS), University of Macau, Macau, China
2. Department of Computer and Information Science, Faculty of Science and Technology, University of Macau, Macau, China

✉ Corresponding author: Xiaoshan Li, Email: xsl@umac.mo

## Abstract

This paper takes up the problem of medical resource sharing through MicroService architecture without compromising patient privacy. To achieve this goal, we suggest refactoring the legacy EHR systems into autonomous MicroServices communicating by the unified techniques such as RESTFul web service. This lets us handle clinical data queries directly and far more efficiently for both internal and external queries. The novelty of the proposed approach lies in avoiding the data *de-identification* process often used as a means of preserving patient privacy. The implemented toolkit combines software engineering technologies such as Java EE, RESTful web services, JSON Web Tokens to allow exchanging medical data in an unidentifiable XML and JSON format as well as restricting users to the *need-to-know* principle. Our technique also inhibits retrospective processing of data such as attacks by an adversary on a medical dataset using advanced computational methods to reveal Protected Health Information (PHI). The approach is validated on an endoscopic reporting application based on openEHR and MST standards. From the usability perspective, the approach can be used to query datasets by clinical researchers, governmental or non-governmental organizations in monitoring health care and medical record services to improve quality of care and treatment.

## Introduction

Patients' Electronic Health Records (EHRs) are stored, processed, and transmitted across several healthcare platforms and among clinical researchers for on-line diagnostic services and other clinical research. This data dissemination serves as a basis for prevention and diagnosis of a disease and other secondary purposes such as health system planning, public health surveillance, and generation of anonymized data for testing. However, exchanging data across organizations is a non-trivial task because of the embodied potential for privacy intrusion. Medical organizations tend to have confidential agreements with patients, which strictly forbid them to disclose any identifiable information of the patients. Health Insurance Portability and Accountability Act (HIPAA) explicitly states the confidentiality protection on health information that any sharable EHRs system must legally comply with. To abide by these strict regulations, data custodians generally use de-identification[1]techniques [1,2] so that any identifiable information on patient's EHR can be suppressed or generalized.

However, in reality, research [3] indicates that 87% of the population of U.S. can be distinguished by sex, date of birth and zip code. We can define quasi-identifiers as the background information about one or more people in the dataset. If an adversary has knowledge of these quasi-identifiers, it can possibly recognize an individual and take advantage of his clinical data. On the other hand, we can find out most of these quasi-identifiers have statistical meanings in clinical research. There exists a paradox between reducing the likelihood of disclosure risk and retaining the data quality. For instance, if information related to patients' residence was excluded from the EHR, it would disable related clinical partners to catch the spread of a disease. Thus, strictly filtered data may lead to failure in operations. Conversely, releasing data including patients' entire information including residence, sex and date of birth would bring a higher disclosure risk.

---

[1] De-identification process is defined as a technology to delete or remove the identifiable information such as name, and SSN from the released information, and suppress or generalize quasi-identifiers, such as zip code date of birth, to ensure that medical data is not re-identifiable (the reverse process of de-identification.)

In this paper we address the emerging problem of de-identification techniques, namely, the problem of offering de-identified dataset for a secondary purpose that makes it possible for a prospective user to perform retrospective processing of medical data endangering patient privacy. Figure 1 overviews the proposed technique, and the standard data request process. Our approach differs from the traditional techniques in the sense that it employs software engineering principles to isolate and develop key requirements of data custodians and requesters. We apply MicroService Architecture [4,5] that provides an effective solution for connecting business functions across the web—both between and within enterprises.

We also present a prototype of our evolving toolset named *MicroShare*, implemented using web services to handle data queries. The results are retrieved in an XML data format that excludes all personal information of patients. The basic model used here follows the principles of RESTful web services by combining three elements: a *URLs repository* for identifying resources uniquely corresponding to clinical data queries, *service consumers* requesting data, and *service producers* as custodians of clinical data. The idea of combining web services with SQL queries is although not new, but it tends to provide a technological approach to avoid medical data re-identification risks. The implemented toolkit uses Java EE that offers an easy way to develop applications using EJBs. Needless to mention that Java EE is widespread and is largely used by community.

Our proof-of-concept implementation uses GastrOS, an openEHR [7] database[2] describing an endoscopic application. The underlying technique provides the ability to construct or use stored queries on a clinical dataset. Employing this clinical toy data warehouse of the GastrOS prototype is a useful way to demonstrate queries on medical data for secondary use. The proposed technique avoids compromising patients' personal information without utilizing de-identification framework tools. For instance, the following query can be posed to GastrOS database

[2] http://gastros.codeplex.com

using our toolkit:

–Find the number of patients who are still susceptible to developing a Hepatitis B infection even after full compliance to the Hepatitis B vaccination schedule–i.e. the baseline and second detection dates for the HBsAg and Anti-HBs tests both show negative results.

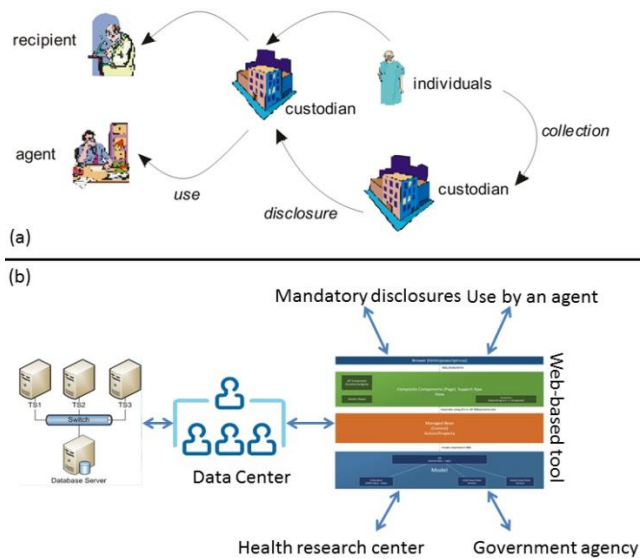

**Figure 1. a)** shows a traditional lifecycle of medical datasets. Custodians can be hospitals, agents may be entities working on their behalf, and recipients are individuals, or organizations such as a pharmaceutical company [6], **b)** depicts the proposed approach that links external entities to data centers using a web interface. The approach excludes all direct data accesses on a dataset.

The set of clinical data queries described in the paper have been crafted with the help of clinical researchers at Vanderbilt University. Supporting such complex queries required developing a set of tools, to which this paper provides the first attempt. In contrast to recent developments on big data, this paper does not focus on the management challenges of medical dataset repositories, but rather focuses on software engineering solutions to deal with the challenges of querying medical data endangering patient privacy. Our approach mainly contributes to the development of privacy preserving techniques on patient data by treating datasets as *blackboxes*. In this way, disclosure risks associated with patient data are minimized. One of the key constraints before accomplishing this goal requires keeping the *computability* with data custodians. Relocating datasets is not only unsafe but leads to data re-identification attempts. To ensure that legitimate users access and execute clinical data queries, we implement an authentication and authorization mechanism using role-based access control (RBAC). RBAC offers a flexible architecture that manages users from different organizations by assigning roles and their corresponding permissions.

***Note that:*** This paper is made substantial extensions to our conference papers [8,9]. We 1) adopt system architecture from monolith Java EE to MicroService for scalability and reliability, and 2) extend EHR resource sharing with additional medical data and queries, 3) support both XML and more readable and efficient JSON for resource format, 4) support accessing control for displaying URL paths and invoke RESTFul web services through JSON Web Token (JWT), 5) reimplement resource queries from structure SQL to object-oriented JPQL, and 6) open source MicroShare at GitHub. The details will be presented in the section 7.

The paper proceeds as follows: Section 2 describes the related work. Section 3 states an application example. Section 4 presents the technical details of our approach. Section 5 overviews the clinical data queries corresponding to the GastrOS dataset. Section 6 discusses the authentication and authorization mechanism connecting users to clinical datasets. Section 7 mentions the extensions from the original papers. Section 8 summarizes the work and details some future research directions.

## Related Work

In contrast to some of the existing techniques [10,11,12,13], our approach relies on advanced software engineering principles and technologies for analyzing clinical datasets. For example, caGrid 1.0 [11] (now caGrid 2.0), released in 2006, is an approach that discusses a complex technical infrastructure for biomedical research through an interconnected network. It aims provide support for discovery, characterization, integrated access, and management of diverse and disparate collections of information sources, analysis methods, and applications in biomedical research. caGrid 1.0 has been initially designed only for cancer research. caGrid combines Grid computing technologies and the Web Services Resource Framework (WSRF) standards to provide a set of core services, toolkits for the development and deployment of new community provided services, and APIs for building client applications. However, caGrid does not focus on an explicit query mechanism to infer details from medical datasets, as the one proposed here. Similar work [14] in discusses a combined interpretation of biological data from various sources. This work, however, considers the problem of continuous updates of both the structure and content of a database and proposes the novel database SYSTOMONAS for systems biology of pseudomonas. Interestingly, this technique combines a data warehouse concept with web services. The data warehouse is supported by traditional ETL (extract, transform, and load) processes and is available at

http://www.systomonas.de.

De-identification techniques for medical data have been studied and developed by statisticians dealing with integrity and confidentiality issues of statistical data. The major techniques used for data de-identification are (i) CAT (Cornell Anonymization Kit) [2], (ii) $\mu$-Argus[3], and (iii) sdcMicro [1]. CAT anonymizes data using generalization, which is proposed [10] as a method that specifically replaces values of quasi-identifiers into value ranges. $\mu$-Argus is an acronym for Anti-Re-Identification General Utility System and is based on a view of safe and unsafe microdata that is used at Statistics Netherlands, which means the rules it applies to protect data comes from practice rather than the precise form of rules. Developed by Statistics Austria, sdcMicro is an extensive system for statistical computing. Like $\mu$-Argus, this tool implements several anonymization methods considering different types of variables. The paper [10] has reported a comparison on the efficacy of these numerical methods that are used to anonymize quasi-identifiers in order to avoid disclosing individual's sensitive information. The Privacy Analytics Risk Assessment Tool (PARAT)[4] is the only commercial product available so far for de-identifying medical data. quantitative analysis [10] of de-identification tools shows that de-identifying data provides no guarantee of anonymity [15]. A study [16] also shows that organizations using data de-identification are vulnerable to re-identification at different rates.

Another approach [17] describes a special query tool developed for the Shared Pathology Informatics Network (SPIN) and integrated into the Indiana Network for Patient care (INPC). This tool allows retrieving de-identified data sets using complex logic and auto-coded final diagnoses, and it supports multiple types of statistical analyses. However, much of the technical details have not been published; for example, the use of complex logic. This and other similar efforts [18] are mostly data-centric. A slightly similar work to this paper has been developed at Massachusetts General Hospital (QPID Inc.[5]), offering solutions at a commercial level, but no prototype is available to experiment with. A Web-based approach for enriching the capabilities of the data-querying system is also developed [12] that considers three important aspects including the interface design used for query formulation, the representation of query results, and the models employed for formulating query criteria. The notion of differential privacy [19] aims to provide means to maximize the accuracy of queries from statistical databases while minimizing the chances of identifying its records.

---

[3] https://github.com/sdcTools
[4] http://www.privacyanalytics.ca/software/
[5] http://www.qpidhealth.com

Our analysis shows that the effort to secure medical datasets is mainly two-faceted: 1) most research endeavors have explored the design and development of de-identification tools, and, 2) some work, mostly led by medical doctors, has tried to address the construction of clinical queries, but they do not provide technical details on the construction of their toolsets. Our approach that treats medical datasets as blackboxes mainly considers the automation of services expected from a data custodian in order to minimize data disclosure risks and making clinical datasets easily accessible for internal and external users.

## GastrOS: An Example Application

GastrOS[6], an openEHR database describing an endoscopic application, is used as a case-study of electronic medical data. This application formed part of the research done at University of Auckland by Koray Atlag in 2010 that investigated software maintainability and interoperability. For this, the domain knowledge model of Archetypes and Templates of openEHR has driven the generation of its graphical user interface. Moreover, the data content depicting the employed terminology, record structure and semantics were based on the Minimal Standard Terminology for Digestive Endoscopy (MST) specified by the World Organization of Digestive Endoscopy (OMED) as its official standard.

Employing the clinical toy data warehouse of the GastrOS prototype is a useful way to demonstrate clinical research-based queries on medical data for secondary use without compromising patients' personal information by using the approach proposed here. The queries shown here focus on endoscopic findings that provide valuable anonymized information to clinicians. The implemented queries are to be mainly used by medical practitioners and health decision-makers alike to help them in their clinical management of patients at the point-of-care and in formulating appropriate health policies, respectively. For example, the following queries are obtained through brainstorming with medical doctors to illustrate our approach.

- Total number of dialysis endoscopic examination from January 1, 2010 to December 31, 2010.

- Top 5 diagnoses for those patients who received endoscopic examination and the number of cases for each diagnosis from January 1, 2010 to December 31, 2010.

- Age profile of endoscopic patients from January 1, 2010 to December 31, 2010? i.e. number of dialysis patients belonging to each of the age bracket [below 18; 18 to below 40; 40 to below 60; 60 and above.

- Number of patients who are still susceptible to developing a Hepatitis B infection even after full compliance to the Hepatitis B vaccination schedule–i.e., the baseline and second detection dates for the HBsAg and Anti-HBs tests both show negative results.

The queries given above are only a subset of original queries. The database structure of GastrOS application is described below.

### GastrOS data structure

Figure 2 describes the data structure of the GastrOS database. GastrOS database contains the following tables: the *clinicaldetection* (doctor detection records), *patient* (patient information), and *examination* (examination records) tables are stored in the database.

The table has two relations: one patient may have more than one clinical detection record or examination record by doctor(s), so the is added as a foreign key in tables. GastrOS is a toy database example with insufficient amount of data available. The original database contains less than 20 rows in each table that makes is not useful for our SQL queries. Therefore, we automatically generated virtual data of 10,000 entries (note that any real data on patients also cannot be published.) An example of the generated data is given in Figure 3. Table 1 provides the up-to-date information on the number of entries in each column of the GastrOS database.

**Table 1.** Generated data in tables

| Table | Row | Size |
|---|---|---|
| ClinicalDetection | 6,393 | 432 KB |
| Examination | 2,020 | 272 KB |
| Patient | 1,881 | 224 KB |
| Sum | 10,294 | 928 KB |

## The Proposed Technique

The proposed approach implements a three-tier application and is devoid of releasing medical datasets, as opposed to traditional techniques. The major purpose and characteristic of the technique extends relatively new software technologies for supporting clinical data queries. To support clinical queries under consideration, we develop an integrated application using SOA and Java EE (Enterprise Edition), to extract data from GastrOS database. There are a plenty of other commercial containers such as JBOSS (Redhat), Websphere (IBM), Weblogic and Glassfish (Oracle), which could be used for our purpose. However, our prototype tool combines Java EE based on JSF Primeface, EJB, and Java Persistence Architecture API (JPA). JPA is a Java specification for accessing, persisting, and managing data between Java objects/classes and a relational database. REST architecture, underlying RESTful web services, treats everything as a resource and is identified by an URI. Resources are handled using POST, GET, PUT, DELETE operations that are identical to Create, Read, Update and Delete (CRUD) operations. Note that in our toolkit it is sufficing to implement operations for handling the described queries. Every request from a client is handled independently, and it must contain all the required information to interpret the request.

---

[6] http://gastros.codeplex.com

## Implementing Clinical Queries using SOA

Web-based authorization and authentication is enforced using role-based access control, before allowing any queries to be accessible by external entities. For instance, the first two queries are shown in Figure 4. They are linked to Organization A, that shows a limited access varying according to the enabled permissions by a security administrator. Listing 1.1 shows the result of applied query. SQL queries, exception results, and running time are presented in columns 1, 2, and 3, respectively.

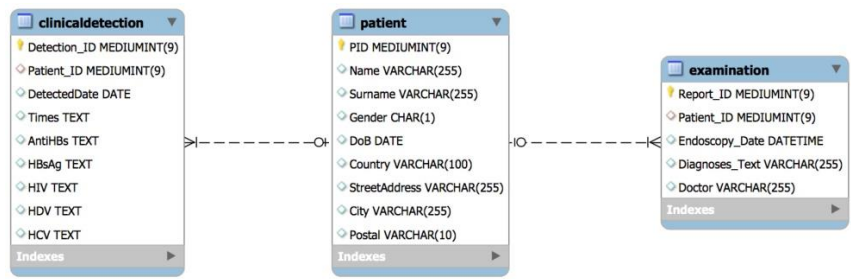

**Figure 2.** E-R Diagram

| PID | Name | Surname | Gender | DoB | Country | StreetAddress | City | Postal |
|-----|------|---------|--------|-----|---------|---------------|------|--------|
| 10000 | Adena | Reeves | F | 1962-08-28 | Montenegro | P.O. Box 936, 9290 Aptent Ave | Morkhoven | 71344 |
| 10001 | Buffy | Warner | M | 2009-05-25 | Guinea-Bissau | P.O. Box 624, 5536 Nunc St. | Graz | 68114-186 |
| 10002 | Kaye | Green | F | 1994-07-23 | Norway | P.O. Box 650, 1264 Tellus. St. | Bojano | T04 3WO |
| 10003 | Keiko | Gonzalez | M | 1973-12-27 | Iraq | 1889 Magna. Street | Chelsea | 8064 |
| 10004 | Kylynn | Carver | F | 1974-01-22 | Tanzania | Ap #357-247 Per Rd. | Oberhausen | 53534 |
| 10005 | Daquan | Sosa | M | 1961-12-28 | Holy See (Vatican City State) | Ap #727-5534 Mauris. Avenue | Eberswalde-Finow | 5690ER |
| 10006 | Rebekah | Navarro | F | 1974-02-01 | Saudi Arabia | P.O. Box 698, 3686 Dui. Avenue | Wolvertem | 2976 |
| 10007 | Zane | Benson | M | 2002-10-19 | Mauritania | Ap #852-3480 Ornare Ave | Dufftown | 1137 |
| 10008 | Jennifer | Petty | F | 1985-08-31 | Isle of Man | 598-2436 Sit Rd. | Bathurst | 71612 |

**Figure 3.** Data generated of patient table

| Id | Description | SQL Query | Service URL | Running Along | Running with Ajax |
|----|-------------|-----------|-------------|---------------|-------------------|
| 2 | Top 5 diagnoses for those patients who received endoscopic examination and the number of cases for each diagnosis from January 1, 2010 to December 31, 2010 | SELECT Diagnoses_Text, COUNT( Diagnoses_Text ) AS Num FROM Examination GROUP BY Diagnoses_Text HAVING COUNT( Diagnoses_Text ) >0 LIMIT 5 | rws/querytwo | RunNoA | RunA |
| 3 | Age profile of endoscopic patients from January 1, 2010 to December 31, 2010 – i.e. Number of dialysis patients belonging to each of the age bracket [below 18; 18 to below 40; 40 to below 60; 60 and above] | SELECT * FROM ( SELECT DISTINCT COUNT( PID ) AS NumBelow18 FROM Patient, Examination WHERE Patient.PID = Examination.Patient_ID AND YEAR( CURRENT_DATE( ) ) - YEAR( DoB ) <18 ) AS NumBelow18, ( SELECT DISTINCT COUNT( PID ) AS Num18to40 FROM Patient, Examination WHERE Patient.PID = Examination.Patient_ID AND YEAR( CURRENT_DATE( ) ) - YEAR( DoB ) BETWEEN 18 AND 40 ) AS Num18to40, ( SELECT DISTINCT COUNT( PID ) AS Num40to60 FROM Patient, Examination WHERE Patient.PID = Examination.Patient_ID AND YEAR( CURRENT_DATE( ) ) - YEAR( DoB ) BETWEEN 40 AND 60 ) AS Num40to60, ( SELECT DISTINCT COUNT( PID ) AS NumAbove60 FROM Patient, Examination WHERE Patient.PID = Examination.Patient_ID AND YEAR( CURRENT_DATE( ) ) - YEAR( DoB ) >60 ) AS NumAbove60 | rws/querythree | RunNoA | RunA |

**Figure 4.** Query list for role of organization A Note that XML-based format is devoid of platform and programming language dependencies. Using this Web-based approach a diverse set of queries can be supported to query clinical data repositories. For the RESTful-based web services, that is the table *urlforwebservice*. A code snippet is given in Listing 1.2 that reveals how the SQL queries are constructed. Note that all the data saved in a program are objects; nonetheless, our database has been represented in the form of relational tables. For this, it needs to implement some ORM (Object-Relational Mapping) techniques. In our prototype implementation we have used JPA (Java Persistence API), because it comes with Java EE technique framework and can be run in either native SQL, or in an object form to allow data manipulation. For instance, we show a *service* code snippet in Listing 1.2. **@Path** show the URL address for this web service, **@GET** is the method of Restful-based web service, that can be used for other reasons such as **@UPDATE @DELETE @POST**. Upon invoking a web service using URL in browser or a session bean, the SQL can be executed and return result by query method which invokes the entity manager of JPA.

```xml
<?xml version="1.0" encoding="utf-8"?>
<dataset>
 <item>
  <element>
   Diagnoses_Text Colon: Primary malignant tumor,
   Quiescent Crohn's disease
  </element>
  <element>421</element>
 </item>
 <item>
  <element>
   Diagnoses_Text Esophagus: Normal, Ectopic gastric mucosa
  </element>
  <element>394</element>
 </item>
 <item>
  <element>Esophagus: Reflux esophagitis</element>
  <element>414</element>
 </item>
 <item>
  <element>Esophagus: Varices certain</element>
  <element>406</element>
 </item>
 <item>
  <element>Esophagus:Barrett's esophagus</element>
  <element>365</element>
 </item>
</dataset>
```

**Figure A. Listing 1.1.** Generated XML Data

XML data retrieved for a couple of queries are shown in Listings 1.1 and 1.3. The corresponding queries are given below, respectively.

- Number of patients for each gender who are still susceptible to developing a Hepatitis B infection even after full compliance to the Hepatitis B vaccination schedule --i.e. the baseline and second detection dates for the HBsAg and Anti-HBs tests both show negative results.

- Top 5 diagnoses for those patients who receive dialysis treatment and the number of cases for each diagnosis from January 1, 2010 to December 31, 2010.

```java
\\For the code of Restful-based web service.
@Path("queryone")
public class QueryOne {
    @Context
    private UriInfo context;
    @EJB
    QueryBean bean;
    @GET
    @Produces("application/xml")
    public String getHtml() {
        // TODO return proper representation object
        String sql = "select Country, COUNT(Report_ID ) AS" +
            "TotalNum" +
            "FROM examination, patient" +
            "WHERE examination.Patient_ID =" +
            "patient.PID" +
            "AND Endoscopy_Date" +
            "BETWEEN \'2010-1-1\'" +
            "AND \'2010-12-30\'" +
            "GROUP BY Country " +
            "Order By TotalNum desc";
        String f = bean.query(sql);
        return f;
    }
}
\\For the method query:
public String query(String sql)
{
    String result = "";
    Query query = emf.createEntityManager().
        createNativeQuery(sql);
    @SuppressWarnings("unchecked")
    List<Object[]> list = query.getResultList();
        ......
}
```

**Figure B. Listing 1.2.** Java code of web service

```
<dataset>
<item>
<element>F</element>
<element>184</element>
</item>
<item>
<element>M</element>
<element>192</element>
</item>
</dataset>
```

**Figure C. Listing 1.3.** Generated XML data

## Enabling dynamic clinical queries

The construction and execution of clinical queries on a given dataset are implemented through a web-interface of the tool. The interface allows a user to dynamically construct a clinical query on a dataset. Thus, it adds a greater flexibility to the query mechanism in developing user-oriented analysis of a dataset. For instance, Figure 5 demonstrates how to execute a query such as "total number of dialysis endoscopic examination of a country starting and ending on a date, respectively.", followed by the output in Figure 6.

These queries show that all specific details on patients are avoided when executing a query, which also means that it disables all direct accesses to patient records. It is actually realized by providing a more aggregated form of data on patients instead of conventional techniques that provide medical datasets to infer such details. Note that the toolkit does not allow any query that provides specific information on patients, such as *"Provide details of all patients with a certain age"*. These queries are directly irrelevant to researchers since they are mainly interested in collective analysis on a dataset. The idea of combining web services with SQL queries is although not new, but it tends to provide a technological solution to a technological problem avoiding medical data re-identification risks. The rationale Using Java EE stems from the fact that it provides a straightforward way to develop applications, for example, EJB are convenient to use by adding only one annotation. Java EE is also widespread being largely used both in academia and industry.

## Authentication and Authorization Process

Our prototype system implements role-based access control [20]. For example, for a medical dataset, operations might include insert, delete, append, and update. The data model of RBAC is based on five data types: users, roles, objects, permissions and executable operations by users on objects.

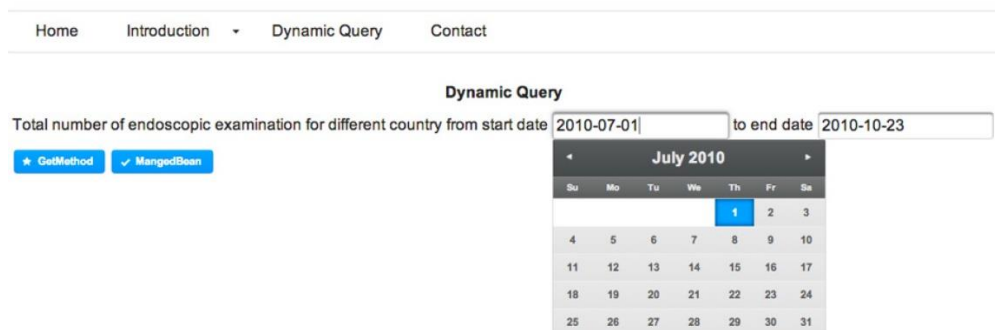

**Figure 5.** Interface for executing runtime clinical queries

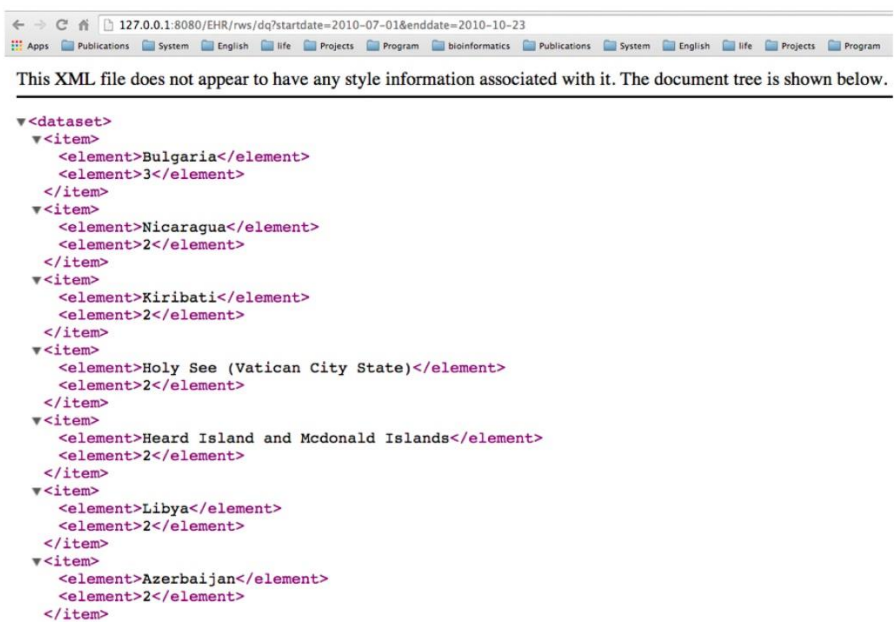

**Figure 6.** The retrieved data in XML format corresponding to the query in Figure 5

A sixth data type, session, is used to associate roles temporarily to users. A role is considered a permanent position in an organization whereas a given user can be switched with another user for that role. Thus, rights are offered to roles instead of users. Roles are assigned to permissions that can later be exercised by users playing these roles. Modeled objects in RBAC are potential resources to protect. Operations are viewed as application-specific user functions. For example, Figure 7 shows a list of queries provided to an administrator role.

To maintain a set of permissions on GastrOS database, we use the constructs from RBAC maintain, and enlist entries in corresponding tables *user*, *roles*, *textsfquerytoroles*, *querylist*, and *urlforwebservice*. We create a user account in user table with the assigned role. Here, all the roles are defined in table. Users privileges and a list of queries are defined in tables *querytorole* and *querylist*, respectively. URLs are stored in the table. For example, logging in as administrator provides five SQL queries shown in Figure 7, whereas logging in as allows a restricted set of SQL queries as given in Figure 4. Security management is supervised by an administrator who can do deletion, addition of roles as required. Using RBAC allows users to take multiple roles, for example, the user could act as researcher that belongs to organization A but can be assigned another role from the set of roles. Similarly, a permission can be associated to many roles depending on the RBAC policy. The multi-to-multi relation between roles and queries that is given in the *querytorole* table (Figure 8).

## Avoiding SQL injections and sensitive information release

Web application security vulnerabilities occur in cases when an attacker or an authorized user tries to submit and execute a database SQL command on a web application, and thus, a back-end database is exposed to an adversary. These SQL injections can be avoided if queries are validated and filtered before their execution and are checked against input data or any encoding made by a user. To prevent similar security issue in our web application we first authenticate the user input against a set of defined rules given below:

$$BlockList = \{name, age, address, zipcode\}$$

$$AntiInjectionList = \{', ", etc. \}$$

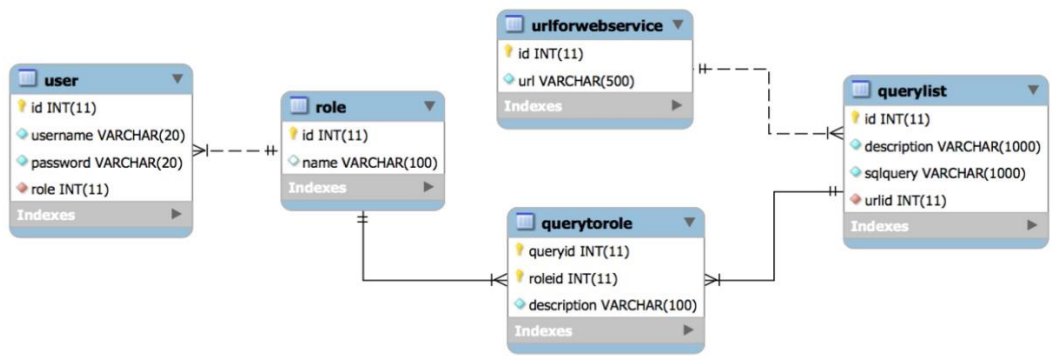

**Figure 7.** Query list for role of administrator

**Figure 8.** E-R Diagram

Note that the special characters given in a block list helps to avoid SQL injections. The set disables all possible access to attributes in a table such as name, age, address, and zip code to keep the fetched data completely anonymized. Set members in filters out three possible vulnerable inputs, i.e., so that any similar attempts could be restricted. Here are the filters that check inputs against. Before running a web service, these two atomic services are always invoked to avoid identifying the actual patients and SQL injections.

Service one: Checks input string for characters in BlockList.

```
bool CheckDeIdentification(String s)
{
    Check Input string s,
    if it contain character in BlockList,
    return false. Otherwise true.
}
```

Service two: Checks input string for characters in AntiInjectionList.

```
bool CheckInjection(String s)
{
    Check Input string s,
    if it contain character in Anti-injectionList,
    return false. Otherwise true.
}
```

## Extension to the Conference Papers

We made some substantial extensions from our conference papers: monolith architecture to MicroService architecture, extending EHR with medical data and queries, support both XML and JSON, resource sharing with JWT, Native SQL to

JPQL, and open source MicroShare at GitHub. The details will be presented in the following subsections.

### Monolith Java EE to MicroService Architecture

EHR system usually adopts the traditional heavy stacks (Figure 9) such as Java EE and .NET Platform and deploying them into enterprise-class application container for achieving high stability and reliability [21]. But the large-size monolith system is difficult to extend and communicate with external system because of the complex dependency and implementation techniques [22]. RESTFul Web Services [23] is a promising technology, which provides standard HTTP wrapper on the internal business layer for the external communications. From this point, our papers [8,9] used RESTFul web services to encapsulate medical resource for external queries. This approach makes sharing medical data from the monolithic system without too much efforts. However, this will lead to a disaster when too many services are required to serve the external queries. We not only need to implement many wrappers for the different services of the business layer, but also face to dependency hell. MicroService architecture [21] can well-handle this issue by decompose the monolithic system into small independently and autonomous components with standard communication protocol like HTTP. This will provide a unified environment for both internal and external communication. Then the original wrapper for external queries does not need to be re-implemented. Moreover, microservice architecture offers the standard authentication mechanism JSON Web Token (JWT) [24] to make flexibly control to the system services, allowing them to be reached from outside customers.

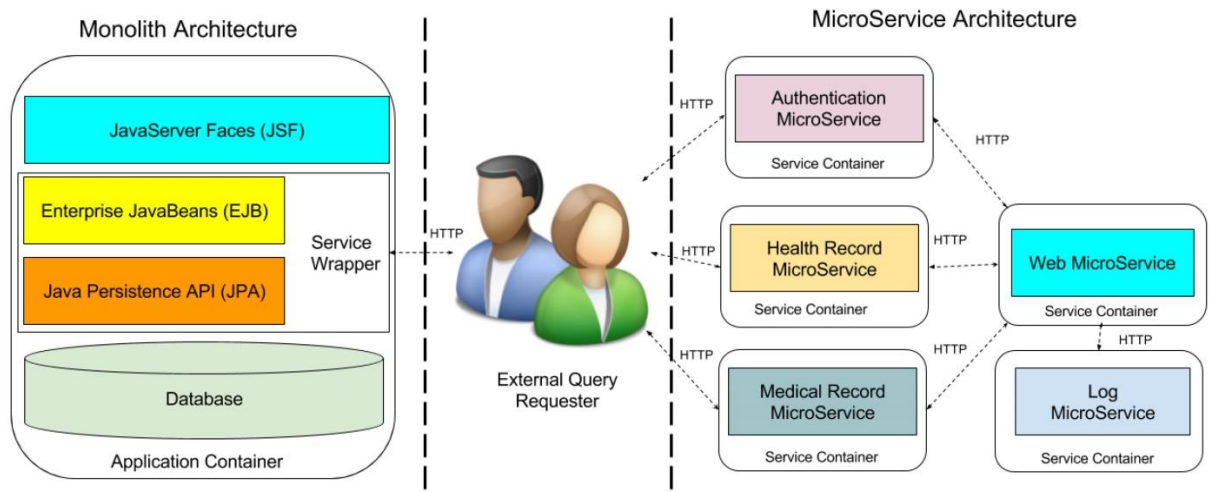

**Figure 9.** Monolith Architecture vs MicroService Architecture. Note that JSF component and Web MicroService are GUIs for display the available RESTFul queries for the requester. Once the requester retrieved the RESTFul path of the queries, the query requests are directly sent to the corresponding sharing MicroService, then they directory return the query result to the requesters.

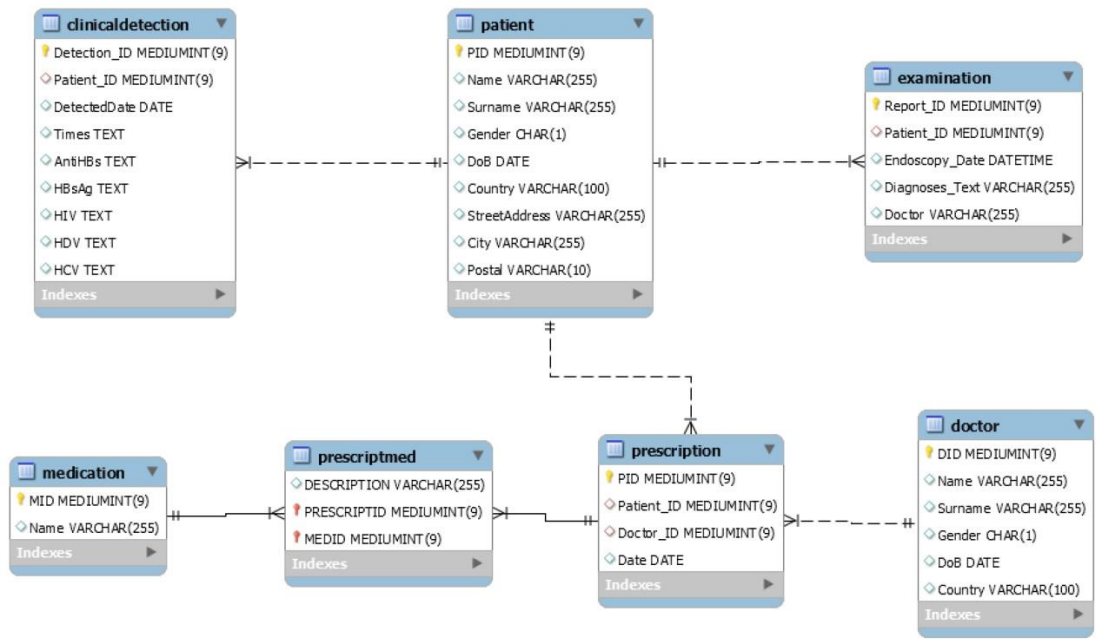

**Figure 10.** E-R Diagram of extending tables

Under the guidelines of microservice profile[7], we decompose our proposed monolith Java EE architecture into five microservice. Left side of Figure 9 contains the monolith architecture in Java EE enterprise stacks. The external query requester is in the middle of the figure. Right side holds the decomposed microservice architecture, which includes web microservice, health record microservice, medical record microservice, authentication microservice, and log microservice. Web microservice provides the graphical user interface to patients and doctors to use the EHR system, it requires the services from health record and medical microservices. Authentication microservice identifies the identification of patients and doctors, and the external query requesters. Furthermore, all above microservices require the audit through log microservice to records all the activities in the system for supporting the post-breach analysis. Comparison to monolith architecture, microservice architecture is more scalable and maintained for sharing any microservices to external requester without implementing the heavy service wrapper. In addition, microservice architecture provides the unified authentication portal for all the internal and external queries and preserves the privacy through additional log microservices. Moreover, each microservice is deployed into independent container such as docker container [25]. That makes microservice architecture cloud-native, that can be directly deployed into cloud platform such as Amazon Web Service, Google Cloud Compute [26], Microsoft Azure [27] or private cloud to reach high reliability and fault tolerance.

## Extending with Medical Data and Queries

To support the medical record microservice, the GastrOS database extends the following tables: *doctor* (doctor information), *prescription* (prescription record which doctors given to patients), *medication* (medication information), and *prescriptmed* (prescription and medication record). Figure 10 describes the extending data structures.

The medical record microservice provides the records of prescriptions which given by doctors to patients, and the records of medications dispensed in prescriptions. In the database, there exists many-to-many relationships between patients and doctors, and between prescription and medication. Thus, the *prescription* table has the foreign keys referring to the tables *patient* and *doctor* and the table *prescriptmed* has the foreign keys referring to the tables *prescription* and *medication*. Table 2 provides the information on the number of entries in each extending table of the GastrOS database.

**Table 2.** Generated data in extending tables

| Table | Row | Size |
|---|---|---|
| doctor | 912 | 80 KB |
| prescription | 3,856 | 160 KB |
| medication | 100 | 16 KB |
| prescriptmed | 8,801 | 4 MB |

---

[7] https://microprofile.io/

Based on the extending data structures, the medical record microservice can provides the medical relevant queries, illustrated by the followings:

- Total number of prescription made up from January 1, 2017 to December 31, 2017.

- Total number of patients who received medical prescription from the doctor Tom Baker from January 1, 2017 to December 31, 2017.

- Age profile of patients who took the medication Abilify from January 1, 2017 to December 31, 2017.

- Top 5 medication given in prescriptions and for each medication the number of patients who took it from January 1, 2017 to December 31, 2017.

## Supporting both XML and JSON Data Format

XML is the standard message format for SOAP-based Web Service [28], it is widely used configuration file format in the software development. However, it is not readable for human and not effective for front-end processing of web application. JavaScript Object Notation (JSON) is an alternatives of XML format for lightweight exchange resources [29]. It is native supported by the front-end programming language JavaScript. JavaScript can load the use JSON without any extra efforts. The corresponding JSON format of XML in Figure 6 and Listing 1.3 are as following:

```
(A)
[
    {
        "DiagnosesText": "Diagnoses_Text Colon: Primary malignant tumor, Quiescent Crohn's disease",
        "Number": "421"
    },
    {
        "DiagnosesText": "Diagnoses_Text Esophagus: Normal, Ectopic gastric mucosa",
        "Number": "394"
    },
    {
        "DiagnosesText": "Esophagus: Reflux esophagitis",
        "Number": "414"
    },
    {
        "DiagnosesText": "Esophagus: Varices certain",
        "Number": "406"
    },
    {
        "DiagnosesText": "Esophagus:Barrett's esophagus",
        "Number": "365"
    }
]
```

```
(B)
[
    {
        "Gender": "F",
        "NumberOfPatients": "184"
    },
    {
        "Gender": "M",
        "NumberOfPatients": "192"
    }
]
```

**Figure 11 A.** JSON Result of Query "Top 5 diagnoses for those patients who receive dialysis treatment and the number of cases for each diagnosis from January 1, 2010 to December 31, 2010." **B.** JSON Result of Query "Number of patients for each gender who are still susceptible to developing a Hepatitis B infection even after full compliance to the Hepatitis B vaccination schedule --i.e. the baseline and second detection dates for the HBsAg and Anti-HBs tests both show negative results."

Figure 11A and 11B shows the query result in JSON format, we can see it much clear and readable than XML. In our proposed MicroShare platform, we support both XML and JSON format, the customer can pick any format as return format as their requirements.

## Query Language from SQL to JPQL

Structured Query Language (SQL) is the standard query language for the relational databases [30]. It is widely used in enterprise application development. However, it is often suffered from the "SQL-injection" attack [31]. The hackers can get the highest privileges of system with the specific designed SQL query. That is why we elaborately build a filter in section 6.1 for the SQL query before submitting them to the SQL engine. There is another obstacle for using SQL in the modern application development. The modern programming language is object-oriented, but the SQL is based on relational database. That mean will need extra works for mapping relational data source to objects. That is the reason using the JPA in Java EE developments. Java Persistence query language (JPQL) [32] is a platform independent object-oriented query language, which can greatly alleviate this issue. The SQL query for "total number of endoscopic examination for different country from January 1, 2010 to December 31, 2010" is in the Listing 1.2, the corresponding JPQL query is as follows:

```
SELECT    new result.dqlistresult(p.country, count(e.reportid))
FROM      examination e,
          patient p
WHERE     e.patientid.pid = p.pid
AND       e.endoscopydate BETWEEN \'2010-1-1\' AND \'2010-12-30\'
GROUP BY p.country
ORDER BY count(e.reportid) DESC
```

We can see the table examination of SQL become to object examination e, the result is columns of tables in SQL become to the properties in the new object dqlistresult. Thanks to JPQL, the developer can stand in the same object-oriented methodology in programming and design the queries. Furthermore, the JPQL can automated filter any harmful character before execution. In short, JPQL can make a query design and execution in to the safe and efficient level.

## Resource Sharing with JSON Web Token

We use role-based access control mechanism to display different queries to different level requesters such as administrator, researchers, and organizations. We know the RESTFul queries is based on the unique Uniform Resource Locator (URL). Although low level customers cannot see the high-level queries, they can access the queries once they can get the URL paths of high level queries. Therefore, we not only need to control the displaying the URLs for different level requesters but also need to control the access to the RESTFul queries. MicroShare adopts industry standard access control methods JSON Web Token (JWT) [24] for representing claims securely among the microservices. Before accessing any RESTFul service of MicroShare, an authentication token must retrieve first from MicroService authentication. Then the external and internal requesters can query the medical resources through the RESTFul sharing service with the retrieved token. For the safety purposes, the token can only use in a limited time with an expired timestamp. When the token is expired, the requesters need to retrieve a new token with their identifications. The detail workflow of resource sharing is shown in Figure 12.

The workflow is: 1) The query requesters get the authentication REST path from web microservices, 2) send authentication request to the authentication microservice to retrieve the token, 3) get the query REST paths from web microservice, 4) send the query request to the health record microservice or medical record microservice with token to retrieve the medical resources.

## Open Source MicroShare

We implemented the proposed approach as the sharing platform named MicroShare. GitHub is the most activity open source community and code repository providers. Unlike the previous paper, we pushed our source code of MicroShare into the GitHub repository[8]. The reader can freely download the source code and make any improvement under the protocol GNU General Public License[9].

## Conclusions and Future Perspectives

We presented a technique for automatic identification of clinically-relevant patterns in medical data. The main contribution of this paper is in defining and presenting an alternative approach to the data de-identification techniques commonly employed for anonymizing clinical datasets. Our technique treats datasets as *blackboxes* and allows data custodians to handle clinical data queries directly. Relocating a dataset not only endangers anonymity of patients, it allows adversaries to apply advanced computational methods for retrospective processing of data. As clinical data is frequently updated, our approach enables data custodians to provide up-to-date resources to their users. We integrate MicroService architecture with a backend clinical database exchanging anonymous JSON and XML data, enabling them to be language and technology independent. Java EE, due to equipped with MicroService architecture, is easy to use for developing applications.

In circumstances related to sharing of patients' data, complex administrative regulations are placed at various levels of management that sometimes unnecessarily complicate the data acquisition process. Providing a tool support for linking data custodians and data requesters using software engineering techniques could pave the way to query clinical datasets more transparently and systematically.

The work provided an initial attempt to build toolset for anonymously analyzing clinical datasets. Our future work includes expanding the approach to more complex databases and supporting an enriched interface for analyzing bigger data repositories. We are currently dealing with the challenge of replacing de-identification techniques in use for de-identifying specific attributes in a database table, for example, patient id, and a doctor needing to find patients who had an increase of systolic blood pressure within a specific period, or patients with steady states of diastolic blood pressure for more than a week. Our future work considers incorporating such queries into the toolset, including implementing ETL processes such as in data warehouses to support clinical data analyses on large-scale integrated databases.

## Acknowledgments

This work was supported by the Macau Science and Technology Development Fund (FDCT) (Grant

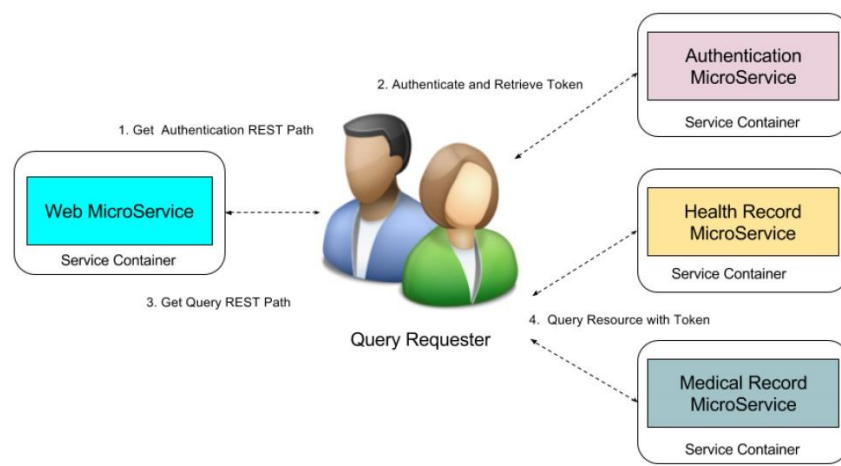

**Figure 12.** MicroShare Resource Sharing Workflow.

No. 103/2015/A3 and 018/2011/A1), Multi-Year Research Grant (MYRG) of University of Macau (Grant No. 2017-00141-FST) and the National Natural Science Foundation of China (NSFC) (Grant No. 61562011).

## Competing Interests

The authors have declared that no competing interest exists.